\documentclass[10 pt]{article}

\usepackage[a4paper, top=2cm, bottom=2cm]{geometry}  
\usepackage[english]{babel}         
\usepackage{graphicx}               
\usepackage{float}                  
\usepackage[colorlinks=true, allcolors=blue]{hyperref}  
\usepackage{amsmath}
\usepackage{mathrsfs}
\usepackage{amsfonts}
\usepackage{csvsimple}
\usepackage{tabularx}
\usepackage{amsthm}
\usepackage{authblk}

\numberwithin{equation}{section}
\graphicspath{ {images/} } 

\newtheorem{theorem}{Theorem}

\newcommand {\beq}{\begin{equation}}
\newcommand {\eeq}{\end{equation}}
\newcommand {\beqa}{\begin{eqnarray}}
\newcommand {\eeqa}{\end{eqnarray}}   
\newcommand {\beqs}{\begin{eqnarray*}}
\newcommand {\eeqs}{\end{eqnarray*}}
\newcommand {\bds}{\begin{displaymath}}
\newcommand {\eds}{\end{displaymath}}
\newcommand {\n}{\nonumber\\}

\newcommand{\lt}{\left}
\newcommand{\rt}{\right}

\nocite{*}

\title{St\"ackel transform, coupling constant metamorphosis and algebraization of quasi-exactly solvable systems}

\author[1]{Siyu Li \footnote{Siyu.Li@latrobe.edu.au}}
\author[2]{Ian Marquette \footnote{i.marquette@latrobe.edu.au}}
\author[3]{Yao-Zhong Zhang \footnote{yzz@maths.uq.edu.au}}
\affil[1]{ Department of Mathematical and Physical Sciences, La Trobe University, Bundoora, VIC 3086, Australia}
\affil[2]{ Department of Mathematical and Physical Sciences, La Trobe University, Bendigo, Victoria 3552, Australia}
\affil[3]{ School of Mathematics and Physics, The University of Queensland, Brisbane, QLD 4072, Australia}

\begin{document}

\maketitle

\begin{abstract}
 \noindent We generalize the notions of the St\"ackel transform and the coupling constant metamorphosis to quasi-exactly solvable systems. We discover that for a variety of one-dimensional and separable multidimensional quasi-exactly solvable systems, their $sl(2)$ algebraizations can only be achieved via coupling constant metamorphosis after appropriate St\"ackel transformations. This discovery has interesting applications, allowing us to derive algebraizations and energies for a wide class of quasi-exactly solvable systems, such as Hooke's atoms in magnetic fields and Newtonian cosmology. The approach of coupling constant metamorphosis was successfully applied previously in the context of exactly solvable, integrable and superintegrable systems. To our knowledge, the present work is the first to apply the idea and approach in the context of quasi-exactly solvable systems.
\end{abstract}

\section {Introduction}
The St\"ackel transform and coupling constant metamorphosis (CCM) have a long history \cite{stackel} that goes back to the study of the Hamilton-Jacobi equation and the Levi-Civita, Kustaanheimo-Stiefel and Hurwitz transformations in the context of the Kepler system and harmonic oscillator in various dimensions \cite{LeviCivita1906, Kustaanheimo1965, 
Hurwitz1923}. These works provided a framework for describing different properties such as regularization and separation of variables. The St\"ackel transform and CCM were applied more systematically from the 1980s \cite{hietarinta1984, boyer1986} in the context of classical integrable systems. They were subsequently extended and applied to the study of quantum integrable systems. In particular, they have been used to define equivalence classes and played an important role in the classification of superintegrable systems \cite{daskaloyannis2008, kalnins2009, miller2013}. 

The St\"ackel transform and CCM are two distinct transformations. The St\"ackel transform consists of using a St\"ackel multiplier while CCM involves a change in the role of the model parameters. However, they are often applied together with additional change of variables to the Schr\"odinger equations. These transforms
preserve the constant of motions and maintain the integrability of both classical and quantum systems \cite{hietarinta1984}. They can also be used to map integrable or superintegrable systems in different manifolds, keeping their integrability or superintegrability property \cite{kalnins2009, kalnins2005, kalnins2006}. 
Two systems are defined as "St\"ackel equivalent" if one system can be mapped to another via a sequence of St\"ackel transforms \cite{boyer1986}. This property is useful in a large number of applications. 
For example, the harmonic oscillator and the Coulomb potential in a $D$-dimensional space were studied and related via the St\"ackel transform and CCM \cite{quesne2016}.  The harmonic oscillator and Kepler-Coloumb potentials in the Euclidean space are transformed by the St\"ackel transform to maximal superintegrable systems in certain Riemannian spaces of nonconstant curvature \cite{ballesteros2011}. 


In this paper, we generalize the notions of St\"ackel transform and CCM widely used in (super)integrable systems \cite{kalnins2009,Post2010} to quasi-exactly solvable (QES) models. In this latter context, we are no longer looking to preserve integrals of motion or separation of variables. Instead, we seek transformations such that a QES system is amenable to an appropriate Lie-algebraic form. 

QES models are quantum mechanical systems that admit only a finite number of eigenvalues and eigenfunctions to be analytically obtained \cite{ushveridze2017, turbiner2016}. The ODEs for QES models are Fuchsian ones of Heun type (see e.g. \cite{zhang2019} and the references therein), and they have applications in a wide range of fields such as condensed matter physics, quantum optics, and black holes. 
It is well-known that Lie algebraization in terms of $sl(2)$ algebra provides a powerful way for identifying and classifying a large class of QES systems \cite{turbiner1988}.
The Lie algebraic framework can be extended to N-body problems via $sl(N+1)$ algebra \cite{hou1999}.  
However, there exist many QES systems of physical interest, whose gauged-transformed Hamiltonians in their original forms do not possess a Lie algebraization directly. In this work, we propose and develop a novel approach involving the St\"ackel transform and CCM to obtain algebraizations for the St\"ckel equivalents of such QES systems.  We will show that even though the initial Hamiltonians of many QES systems do not possess a hidden $sl(2)$ symmetry, their $sl(2)$ Lie algebraizations can be achieved via the St\"ackel transform and CCM. We also obtain closed-form expressions for the wavefunctions, energies, and parameter constraints of these models.  

The paper is organized as follows. In Section 2, we present our general procedure. We first give a proposition showing how to map a Hamiltonian ${\cal H}$ to an equivalent Hamiltonian ${\cal H}'$ via the St\"ackel transform and CCM.  In many practical applications, the St\"ackel transformed Hamiltonian ${\cal H}'$ becomes a Fuchsian (or Heun-type) differential operator involving polynomials of degrees 4, 3, and 2. We then show that if the coefficients of the polynomials satisfy certain algebraic relations, the Heun-type differential operator ${\cal H}'$ is QES and possesses a $sl(2)$ algebraization. This in turn provides a way to determine the constraints of the model parameters, energy spectrum, and wavefunctions of the system. In Sections 3-8, we apply the general procedure to obtain the $sl(2)$ algebraizations for large classes of systems, which otherwise do not seem to be algebraical directly. The wide range of cases also demonstrates the usefulness of the St\"ackel transform and CCM in QES models. 
Systems considered in this paper are: (i) 2D hydrogen atom in a uniform magnetic field, (ii) Two electrons in an external oscillator potential, (iii) Two planar charged particles in a uniform magnetic field, (iv) Two Coulombically repelling electrons on a sphere, (v)  Inverse quartic power potential, (vi) Inverse sextic power potential, and (vii) Newtonian cosmology model.  For all these models, $sl(2)$ algebraizations are obtained by the St\"ackel transforms. We also obtain the constraints of the model parameters, energies, and wavefunctions of the QES models from the St\"akel transformed Hamiltonians via the method of CCM. We conclude the paper with a short summary of our results in Section 9.

\section{St\"ackel transform and algebraization of QES systems}

The Hamiltonian of a quantum mechanical system usually contains many model parameters.   
Only systems whose model parameters satisfy certain constraints are QES. The exact (i.e. closed-form) expressions for the energies and wavefunctions of QES systems can be obtained by solving the corresponding Schr\"odinger equation by means of, e.g., the Bethe ansatz method \cite{zhang2012}. However, in many cases, the Hamiltonian of a QES model in its original form is not Lie algebraic, and Lie algebraization can only be achieved for their St\"ackel equivalent (or dual) system with the help of CCM. 

In this section, we generalize the notions of St\"ackel transforms and CCM, and present a general framework for obtaining Lie algebraization of a QES system whose original Hamiltonian is non-Lie algebraic. 


Let $\mathcal{H}=H(x,p)-\alpha U(x)$ be a certain gauge-transformed Hamiltonian of a quantum mechanical system, where $H(x,p)$ is independent of the parameter $\alpha$ and $U(x)$ is the potential. The time-independent Schr\"odinger equation takes the form
\begin{equation}\label{hamiltonian Schrodinger equation}
    \mathcal{H} \psi(x)=[H(x,p)-\alpha U(x)]\psi(x)=E\psi(x).
\end{equation}
Then we have
\vskip.1in
\noindent {\bf Proposition:} 
{\em   Let $U(x)\neq0$ be a St\"ackel multiplier.  Then the St\"ackel transformed Hamiltonian
\begin{equation}\label{Stackel transformation Hamiltonian}
    \mathcal{H'}=U^{-1}(x)[H(x,p)-\alpha'],
\end{equation}
 describe a quantum mechanical system with the Schr\"odinger equation 
\begin{equation}\label{Stackel transformation Schrodinger equation}
    \mathcal{H'}=U^{-1}(x)[H(x,p)-\alpha']\psi(x)=E'\psi(x).
\end{equation}
This system is equivalent to (or dual to) the system described by $\mathcal{H}$ under the coupling constant metamorphosis:
\begin{equation}\label{dual to}
    \alpha'\iff E,\quad E'\iff \alpha,
\end{equation}
The St\"ackel transformed Hamiltonian $\mathcal{H}'$ is St\"ackel equivalent to the gauge-transformed Hamiltonian $\mathcal{H}$. This implies that the eigenfunctions of $\mathcal{H}'$ is also the eigenfunctions of $\mathcal{H}$.
}

The St\"ackel transform and metamorphosis of the model parameters are the key for the Lie algebraizations of the QES systems. 
In most applications, the St\"ackel transformed Hamiltonian $\mathcal{H}'$ is a differential operator in a single variable $x$ of the following form
\begin{equation}\label{ODE Heun differential equation}
\mathcal{H}'=X(x) \frac{d^2}{dx^2}+Y(x) \frac{d}{dx}+Z(x),
\end{equation}
where $X(x), Y(x)$ and $Z(x)$ are polynomials of degrees 4,3 and 2 respectively,
\begin{align}\label{X Y Z}
    X(x)=\sum_{k=0}^4 a_k x^k, \qquad Y(x)=\sum_{k=0}^3 b_k x^k,\qquad Z(x)=\sum_{k=0}^2 c_k x^k.
\end{align}
Here $a_i, ~b_i$ and $c_i$ are certain constant coefficients related to the model parameters of the St\'ackel transformed Hamiltonian $\mathcal{H}'$.
It can be shown \cite{zhang2016} that differential operator $\mathcal{H}'$ of the above form is $sl(2)$ algebraic if some of the coefficients in the polynomials $X(x)$, $Y(x)$ and $Z(x)$ satisfy certain relations. We have,
\begin{theorem}\cite{zhang2016}\label{theorem sl(2) parameter constrain}
The differential operator $\mathcal{H}'$ allows for a $sl(2)$ algebraization, i.e.
has a hidden $sl(2)$ algebraic structure, if and only if
\beq
b_3=-2(n-1)a_4,\qquad c_1=-n[(n-1)a_3+b_2],\qquad c_2=n(n-1)a_4,
\eeq
where $n$ is a nonnegative integer.
\end{theorem}
Indeed, under these conditions ${\cal H}'$ can be expressed as
\beqa
{\cal H}'&=&a_4J^+J^+-a_3J^+J^0+a_2J^0J^0+a_1J^0J^-+a_0J^-J^--\left(\frac{3n-2}{2}a_3+b_2\right)J^+\n
& &+[(n-1)a_2+b_1]J^0+\left(\frac{n}{2}a_1+b_0\right)J^-+\frac{n}{2}\left[\left(\frac{n}{2}-1\right)a_2+b_1\right]
   +c_0,\label{Suff-H2}
\eeqa
in terms of the differential operators 
\begin{align}
    \label{sl(2) generators}
        J^+=-x^2 \frac{d}{dx}+nx,\qquad J^0=x\frac{d}{dx}-\frac{n}{2},\qquad J^-=\frac{d}{dx}.
\end{align}
These differential operators satisfy the $sl(2)$ commutation relations,
\begin{align}
        \left[J^0,J^{\pm}\right]=\pm J^{\pm},\qquad \left[J^+,J^-\right]=2 J^0.
\end{align}
If $n$ is a nonnegative integer, $n=0,1,2,\cdots,$ then (\ref{sl(2) generators}) provides a $(n+1)$-dimensional irreducible
representation ${\cal P}_{n+1}(x)={\rm span}\{1,x,x^2,\cdots,x^n\}$ of the $sl(2)$ algebra.
It is evident that any differential operator which is a polynomial of $sl(2)$ generators (\ref{sl(2) generators}) with $n$ positive
integer will have the space ${\cal P}_{n+1}(x)$ as its invariant subspace, i.e. have $(n+1)$ eigenfunctions in
the form of polynomials in $x$ of degree $n$.
Note that the QES operator $\mathcal{H}'$ is an element of the universal enveloping algebra $\mathcal{U}[sl(2)]$ of $sl(2)$. 
This is the main idea underlying the Lie algebraic approach to QES problems.


The first $(n+1)$ eigenvalues of $\mathcal{H}'$ can be obtained from the eigenvalues of the Jacobi matrix with the following elements,
\begin{eqnarray}
 &&\mathcal{H}'_{k-2,k}=k(k-1)a_0, \qquad \mathcal{H}'_{k-1,k}=k \left[(k-1)a_1+b_0\right],\nonumber\\
 &&\mathcal{H}'_{k,k}=c_0+kb_1+k(k-1)a_2,\nonumber\\
 &&\mathcal{H}'_{k+1,k}=(k-n)\left[(n+k-1)a_3+b_2\right], \qquad \mathcal{H}'_{k+2,k}=(n-k)(n-k-1)a_4. 
\end{eqnarray}
Obviously, when $k=n$, we have $\mathcal{H}'_{k+1,k}=0=\mathcal{H}'_{k+2,k}$ and thus the Jacobi matrix is a $(n+1)\times (n+1)$ tridiagonal matrix, as expected from the fact that ${\cal H}'$ preserves the $(n+1)$-dimensional polynomial space ${\cal P}_{n+1}(x)$.
Due to the equivalence of St\"ackel, if the St"ackel transformed Hamiltonian $\mathcal{H}'$ is QES, the gauged transformed Hamiltonian $\mathcal{H}$ is also QES and its eigenvalues can be obtained via CCM.

In the next sections, we will apply the above general procedure to a large class of QES models that are not associated directly with a hidden $sl(2)$ algebraic structure. We perform the appropriate St\"ackel transform for each case and obtain the $sl(2)$ algebraizations for the St\"ackel transformed (or equivalent)  systems. This in turn gives the hidden $sl(2)$ algebraic structures of the original systems via CCM. Low-lying energies and the corresponding analytic wave functions of these models are also presented. This illustrates the wide applicability of our method.

\section{2D hydrogen atom in a uniform magnetic field}

Two-dimensional hydrogen atom (or hydrogen-like atom) in a uniform magnetic field has been widely studied in the literature.  
Analytic solutions to this system were first derived in \cite{taut1995}.
We will show that this system is quasi-exactly solvable by establishing the hidden $sl(2)$ symmetry of the St\"ackel equivalent system. This also allows us to obtain the analytical solutions of the orignal model via CCM.

The Hamiltonian of 2D hydrogen in a uniform magnetic field reads
\cite{taut1995,naber2019}

\beq
H_0=\frac{1}{2}\lt({\bf p}+\frac{1}{c}{\bf A}\rt)^2+\frac{Z}{r},
\eeq
where $c$ is the velocity of light and the vector potential in the symmetric gauge is given by ${\bf A}=\frac{1}{2}{\bf B}\times {\bf r}$. The magnetic field ${\bf B}$ is perpendicular to the plane in which the electron is located. In polar coordinates $(r,\theta)$ within the plane, the angular and radial part of the wavefunction $\phi({\bf r})$ are decoupled through the following factorized form of the wavefunctions
\beq\label{2D hydrogen phi(r)}
\phi({\bf r})=\frac{e^{im\theta}}{\sqrt{2\pi}}\frac{u(r)}{\sqrt{r}},~~~~m=0,\pm 1,\pm 2,\cdots \quad .
\eeq
The radial component of the wavefunction $u(r)$ satisfies the radial Schr\"odinger equation
\beq\label{2D hydrogen radial Schrodinger equation}
\lt[-\frac{d^2}{dr^2}+\frac{m^2-1/4}{r^2}+\omega_L^2 r^2+\frac{Z}{r}\rt]u(r)=2(E-m\omega_L)u(r),
\eeq
where $\omega_L=\frac{1}{2}\omega_c=B/2c$ is the Larmor frequency. Applying the following transformation
\beq\label{2D hydrogen u(r)}
u(r)=r^{|m|+1/2}e^{-\frac{\omega_L}{2}r^2} y(r),
\eeq
and substituting (\ref{2D hydrogen u(r)}) into (\ref{2D hydrogen radial Schrodinger equation}), we get for the variable $y$ the ODE
\beq
\frac{d^2 y}{dr^2}+\left(\frac{2|m|+1}{r}-2\omega_L r\right)\frac{dy}{dr}+\left\{[\epsilon-2(|m|+1)]\omega_L-\frac{Z}{r}\right\}y=0. 
\eeq
Using $\epsilon=\frac{2E}{\omega_L}-2m$, we can write the radial Schr\"odinger equation in the form
\beq
{\cal H}y\equiv \lt(H-\frac{\alpha}{r}\rt)y={\cal E}y,
\eeq
where $\alpha=Z,~{\cal E}=[2(|m|+1)-\epsilon]\omega_L$ with $\epsilon=\frac{2E}{\omega_L}-2m$, and 
\beq
H=\frac{d^2}{dr^2}+\lt(-2\omega_L r+\frac{2|m|+1}{r}\rt)\frac{d}{dr}.
\eeq
Applying the St\"ackel transform, we have
\beqa
&&{\cal H}'y=Z y,\nonumber\\
&&{\cal H}'=r(H-{\cal E})=r\frac{d^2}{dr^2}+[-2\omega_L r^2+2|m|+1]\frac{d}{dr}+[\epsilon-2(|m|+1)]\omega_L r.
\eeqa
${\cal H}'$ allows for an $sl(2)$ algebraization if
\beq
[\epsilon-2(|m|+1)]\omega_L\equiv c_1=-n[(n-1)a_3+b_2]\equiv 2\omega_L n,
\eeq
which gives the result in \cite{taut1995}
\beq
\epsilon=2(n+|m|+1),~~~~n=0,1,2,\cdots \quad .
\eeq
Indeed, for such $\epsilon$ values, ${\cal H}'$ is dependent on integer parameter $n$ and can be expressed in terms of the $sl(2)$ generators as
\beq
{\cal H}'=J^0J^-+2\omega_L J^++\lt[\frac{n}{2}+2|m|+1\rt]J^-.
\eeq

The Jacobi matrix representation of the $\mathcal{H}'$ is 
\beq
\mathcal{H}'_{k-1,k}=k(k+2|m|),\quad \mathcal{H}'_{k,k}=0,\quad \mathcal{H}'_{k+1,k}=2\omega_L(n-k).
\eeq
The Jacobi matrix of $\mathcal{H}'$ is a $(n+1)\times (n+1)$ tri-diagonal matrix, as expected, because $\mathcal{H}'_{k+1,k}=0$ when $k=n$. As examples, we consider the $n=1, 2$ cases and obtain the explicit expressions for the corresponding eigenvalues and eigenfunctions of $\mathcal{H}$. 

For $n=1$, from $\mathcal{H}'y=Zy$ we obtain two $Z$ values together with two solutions for $y$,
\begin{align}
    &\begin{aligned}\label{2D hydrogen n=1 eigenvalues}
        Z=\pm \sqrt{2(2|m|+1)\omega_L},
    \end{aligned}\\
    &\begin{aligned}\label{2D hydrogen n=1 eigenfunctions}
        y=r\pm \sqrt{\frac{2|m|+1}{2\omega_L}}. 
    \end{aligned}
\end{align}
By CCM, the relations (\ref{2D hydrogen n=1 eigenvalues}) above from the St\"ackel equivalent $\mathcal{H}'$ provide the constraints of the model parameters of the original system $\mathcal{H}$.
Substituting (\ref{2D hydrogen n=1 eigenfunctions}) into (\ref{2D hydrogen u(r)}) and (\ref{2D hydrogen phi(r)}), we get the energy and wavefunctions of the initial system for $n=1$ (corresponding to the two different values of the model parameter $Z$)
\beq
\begin{split}
    &E_1=(m+|m|+2)\omega_L,\\
    &\phi_{1\pm}({\bf r})=\frac{r^{|m|}}{\sqrt{2\pi}}\exp\left(im\theta-\frac{\omega_L}{2}r^2\right)\left(r\pm \sqrt{\frac{2|m|+1}{2\omega_L}}\right),~~~~m=0,\pm 1,\pm 2,\cdots \quad .
\end{split}
\eeq
For $n=2$, we have 
\begin{align}
    &\begin{aligned}\label{Z-constraints for n=2}
        Z_1=0,\qquad Z_{2,3}=\pm 2\sqrt{(3+4|m|)\omega_L},
    \end{aligned}\\
    &\begin{aligned}
        y_1=r^2-\frac{1+|m|}{\omega},\qquad y_{2,3}=r^2 \pm \sqrt{\frac{3+4|m|}{\omega}} r+\frac{1+2|m|}{2 \omega}. 
    \end{aligned}
\end{align}
The corresponding energy and the wavefunctions for the initial system ${\cal H}$ are
\begin{align}
    &\begin{aligned}
        E_2=(m+|m|+3)\omega_L,
    \end{aligned}\\
    &\begin{aligned}
        &\phi_{2(1)}({\bf r})=\frac{r^{|m|}}{\sqrt{2\pi}}\exp\left(im\theta-\frac{\omega_L}{2}r^2\right)\left(r^2-\frac{1+|m|}{\omega}\right),\\
       &\phi_{2(2,3)}({\bf r})=\frac{r^{|m|}}{\sqrt{2\pi}}\exp\left(im\theta-\frac{\omega_L}{2}r^2\right)\left(r^2 \pm \sqrt{\frac{3+4|m|}{\omega}} r+\frac{1+2|m|}{2 \omega}\right),~~~~m=0,\pm 1,\pm 2,\cdots 
    \end{aligned}
\end{align}
provided that the model parameters satisfy the constraints (\ref{Z-constraints for n=2}).
However, note that from physical perspectives the $Z=0$ constraint is unwanted,  and thus there are only two physically interesting wave functions $\phi_{2(2,3)}({\bf r})$ for $n=2$. 

\section{Hooke-type models of two charged particles}

In this section, we examine Hooke-type models of two charged particles with Coulomb interaction and in external uniform magnetic field. Such models have interesting applications in nuclear, atomic, and solid-state physics. 
Similarly to the $2D$-hydrogen (hydrogen-like) atom systems, Hooke-type atom systems are not directly associated with any $sl(2)$ algebraizations. In this section, we apply the St\"ackel transform and CCM to find their hidden $sl(2)$ symmetry and analytical solutions. 

\subsection{Two electrons in an external oscillator potential}

Consider the model of two interacting electrons in an external oscillator potential. The Hamiltonian of the system reads 
\cite{taut1993,turbiner1994}

\begin{equation}
    H_0=\sum_{i=1}^2\frac{1}{2}\lt({\bf p}_i^2+\omega^2{\bf r}_i^2\rt)+\frac{Z}{||{\bf r}_1-{\bf r}_2||}.
\end{equation}
Using the relative coordinate ${\bf r}={\bf r}_1-{\bf r}_2$ and the center of mass coordinate
${\bf R}=\frac{1}{2}({\bf r}_1+{\bf r}_2)$, which give rise to new momentum operators
\begin{equation}
{\bf p}=-i\nabla_r=\frac{1}{2}({\bf p}_2-{\bf p}_1),~~~~~{\bf P}=-i\nabla_R={\bf p}_2+{\bf p}_1.
\end{equation}
Then the Hamiltonian can be written as
\begin{equation}
H_0=2\lt[\frac{1}{2}{\bf p}^2+\frac{1}{2}\omega_r^2r^2+\frac{Z}{2r}\rt]
    +\frac{1}{2}\lt[\frac{1}{2}{\bf P}^2+\frac{1}{2}\omega_R^2R^2\rt]\equiv H_r+H_R,
\end{equation}
where $\omega_R=2\omega$ and $\omega_r=\frac{1}{2}\omega$.
The total wave function factorizes
\begin{equation}
\psi(1,2)=\phi({\bf r})\xi({\bf R})
\end{equation}
and the Schr\"odinger equation $H_0\psi=E\psi$ separates into
\begin{equation}
H_r\phi({\bf r})=\epsilon \phi({\bf r}),~~~~~H_R\xi({\bf R})=\eta\,\xi({\bf R}),
\end{equation}
with $E=\epsilon+\eta$ being the total energy of the system.

Introduce the spherical coordinates which separate the modulus $r$ from the angular coordinates, giving rise to the ansatz:
\beq
\phi({\bf r})=\frac{u(r)}{r}Y_{lm}(\hat{\bf r}),\qquad \hat{\bf r}={\bf r}/r,
\eeq
where $Y_{lm}(\hat{\bf r})$ is the spherical harmonics. Then the radial Schr\"odinger equation is given by
\beq\label{Hooke atom oscillator radial Schrodinger equation}
\lt(-\frac{d^2}{dr^2}+\omega_r^2 r^2+\frac{Z}{r}+\frac{l(l+1)}{r^2}\rt)u(r)=\epsilon u(r).
\eeq
Making the gauge transformation
\beq\label{Hooke atom oscillator u(r)}
u(r)=r^{l+1}e^{-\frac{\omega_r}{2}r^2}y(r)
\eeq
and substituting (\ref{Hooke atom oscillator u(r)}) into (\ref{Hooke atom oscillator radial Schrodinger equation}) we obtain
\beq
\frac{d^2 y}{dr^2}+\left(\frac{2l+2}{r}-2\omega_L r\right)\frac{dy}{dr}+\left[\epsilon-(3+2l)\omega_L-\frac{Z}{r}\right]y=0. 
\eeq
We write the radial Schr\"odinger equation in the form
\beq
{\cal H}y\equiv \lt(H-\frac{\alpha}{r}\rt)y={\cal E} y
\eeq
with $\alpha=Z$, ${\cal E}=(3+2l)\omega_L-\epsilon$ and
\beq
H=\frac{d^2}{dr^2}+\lt(-2\omega_r r+\frac{2(l+1)}{r}\rt)\frac{d}{dr}.
\eeq

Applying the St\"ackel transform, we get
\beqa
&&{\cal H}'y=Z y,\nonumber\\
&&{\cal H}'=r(H-{\cal E})=r\frac{d^2}{dr^2}+\lt[-2\omega_r r^2+2(l+1)\rt]\frac{d}{dr}+[\epsilon-(2l+3)\omega_r]r.
\eeqa
${\cal H}'$ allows for an $sl(2)$ algebraization if
\beq
\epsilon-(2l+3)\omega_r\equiv c_1=-n[(n-1)a_3+b_2]\equiv 2\omega_r n
\eeq
which gives the energies obtained in \cite{taut1993,turbiner1994}
\beq
\epsilon=(2n+2l+3)\omega_r,~~~~~n=0,1,2,\cdots.
\eeq
Indeed, for such $\epsilon$ values, ${\cal H}'$ is dependent on integer parameter $n$ and can be expressed
in terms of the $sl(2)$ generators as
\beq
{\cal H}'=J^0J^-+2\omega_r J^++\lt[\frac{n}{2}+2(l+1)\rt]J^-
\eeq
So for fixed $n$ (i.e. fixed energy), there are $(n+1)$ solutions to model parameter $Z$ corresponding to $(n+1)$ eigenfunctions. We remark that the hidden $sl(2)$ symmetry of this model was first noted in \cite{turbiner1994} (see also \cite{chiang2001}) without the application of the St\"ackel transform. 

We can solve the Scr\"odinger equation $\mathcal{H}' y=Z y$ for the St\"ackel equivalent system to obtain $Z$ and the corresponding eigenfunctions. The we can perform CCM and obtain the energies and wavefunctions of the original system.  We illustrate this by presenting the explicit expressions for $n=1,2$ cases below.

For $n=1$, we have
\begin{align}
    Z=\pm 2\sqrt{(l+1)\omega_r},\qquad
    y=r\pm \sqrt{\frac{l+1}{\omega_r}}. 
\end{align}
By CCM, we obtain the energy and the wavefunctions of the initial Schr\"odinger equation
\begin{align}
    &\begin{aligned}
        \epsilon_1=(5+2l)\omega_r,
    \end{aligned}\\
    &\begin{aligned}
        u_{1\pm}(r)=r^{l+1} \exp\left[-\frac{\omega_r}{2}r^2\right] \left(r\pm \sqrt{\frac{l+1}{\omega_r}}\right),
    \end{aligned}
\end{align}
corresponding the two values of the model parameter $Z$. 
For $n=2$, we have
\begin{align}
    &\begin{aligned}
        Z_1=0, \quad Z_{2,3}=\pm 2\sqrt{(5+4l) \omega_r}. 
    \end{aligned}\\
    &\begin{aligned}
        y_1=r^2-\frac{3+2l}{2\omega},\quad y_{2,3}=r^2\pm \sqrt{\frac{5+4l}{\omega_r}} r+\frac{l+1}{\omega_r}.  
    \end{aligned}
\end{align}
The energy and wavefunctions of the initial Schr\"odinger equation are obtained by CCM as 
\begin{align}
    &\begin{aligned}
        \epsilon_2=(7+2l)\omega_r,
    \end{aligned}\\
    &\begin{aligned}
        &u_{2(1)}(r)=r^{l+1} \exp\left[-\frac{\omega_r}{2}r^2\right]\left(r^2-\frac{3+2l}{2\omega}\right),\\
        &u_{2(2,3)}(r)=r^{l+1} \exp\left[-\frac{\omega_r}{2}r^2\right] \left(r^2\pm \sqrt{\frac{5+4l}{\omega_r}} r+\frac{1+l}{\omega_r}\right), 
    \end{aligned}
\end{align}
corresponding to the three values of the model parameter $Z$ given above, respectively.
Note that from physical perspective the model with $Z=0$ (i.e. zero Coulomb interaction) is not interesting. However, mathematically, we have three different wavefunctions for $n=2$, as required by the $sl(2)$ symmetry of the system.

\subsection{Two planar charged particles in uniform magnetic field}

Hooke-type models in an external uniform magnetic field have applications in many fields (e.g. quantum dots) and have been studied numerically by approximation methods, such as Hatree-Fock approximation \cite{pfannkuche1993, kumar1990} and WKB approximation \cite{klama1998}. Some analytic solutions of this model were derived in \cite{taut1994} (see also \cite{zhu2005}). In this section, we will consider a system of two planar charged particles in a uniform magnetic field interacting through the combined Coulomb and harmonic potentials. We will find the hidden $sl(2)$ symmetry of the model by means of the St\"ackel transform and obtain its exact soultions by CCM.

The hamiltonian of the system is given by
\beq
H_0=\sum_{i=1}^2\lt[\frac{1}{2}\lt({\bf p}_i+\frac{1}{c}{\bf A}({\bf r}_i)\rt)^2
   +\frac{1}{2}\omega_0^2{\bf r}_i^2\rt]+\frac{Z}{||{\bf r}_1-{\bf r}_2||}
\eeq
where $c$ is the speed of light and ${\bf A}({\bf r}_i)=\frac{1}{2}{\bf B}\times {\bf r}_i$.
Introduce relative and center of mass coordinates ${\bf r}={\bf r}_1-{\bf r}_2$  and
${\bf R}=\frac{1}{2}({\bf r}_1+{\bf r}_2)$, respectively, then the hamiltonian becomes
\beqa
H_0&=&2\lt[\frac{1}{2}\lt({\bf p}+\frac{1}{c}{\bf A}_r\rt)^2+\frac{1}{2}\omega_r^2 r^2+\frac{Z}{r}\rt]
+\frac{1}{2}\lt[\frac{1}{2}\lt({\bf P}+\frac{1}{c}{\bf A}_R\rt)^2+\frac{1}{2}\omega_R^2 R^2\rt]\nonumber\\
&\equiv& H_r+H_R,
\eeqa
where $\omega_r=\frac{1}{2}\omega_0$, $\omega_R=2\omega_0$ and
\beqa
&&{\bf p}=-i\nabla_r=\frac{1}{2}({\bf p}_2-{\bf p}_1),~~~~~{\bf P}=-i\nabla_R={\bf p}_2+{\bf p}_1,\nonumber\\
&&{\bf A}_r=\frac{1}{2}{\bf A}({\bf r})=\frac{1}{2}[{\bf A}({\bf r}_2)-{\bf A}({\bf r}_1)],~~~~~
   {\bf A}_R=2{\bf A}({\bf R})={\bf A}({\bf r}_2)+{\bf A}({\bf r}_1)].
\eeqa
The total wavefunction factorizes
\beq
\psi(1,2)=\xi({\bf R})\phi({\bf r})
\eeq
and the Schr\"odinger equation $H_0\psi=E\psi$ separates into
\beq
H_r\phi({\bf r})=\epsilon \phi({\bf r}),~~~~~H_R\xi({\bf R})=\eta\xi({\bf R})
\eeq
with $E=\epsilon+\eta$ and the following ansatz for the relative motion
\beq
\phi({\bf r})=\frac{e^{i m\theta}}{\sqrt{2\pi}}\frac{u(r)}{\sqrt{r}},~~~~m=0,\pm 1,\pm 2,\cdots
\eeq
The radial wavefunction $u(r)$ satisfies the radial Schr\"odinger equation
\beq\label{Hooke atom magnetic field radial Schrodinger equation}
\lt[-\frac{d^2}{dr^2}+\frac{m^2-1/4}{r^2}+\tilde{\omega}_r^2 r^2+\frac{Z}{r}\rt]u(r)=(\epsilon-m\omega_L)u(r),
\eeq
where $\omega_L=B/2c$ and $\tilde{\omega}_r=\frac{1}{2}\sqrt{\omega_L^2+\omega_0^2}$
is the effective frequency.

The remaining analysis is quite similar to that in the last section for the 2D hydrogen in a magnetic field. Setting
\beq\label{Hooke atom magnetic field u(r)}
u(r)=r^{|m|+1/2}e^{-\frac{\tilde{\omega}_r}{2}r^2} y(r)
\eeq
and substituting (\ref{Hooke atom magnetic field u(r)}) into (\ref{Hooke atom magnetic field radial Schrodinger equation}), we get 
\beq
\frac{d^2 y}{dr^2}+\left(\frac{2|m|+1}{r}-2\tilde{\omega}_r r\right)\frac{dy}{dr}+\left[\epsilon-m\omega_L-2(|m|+1)\tilde{\omega}_r-\frac{Z}{r}\right]y=0. 
\eeq
This ODE can be written in the form
\beq
{\cal H}y\equiv \lt(H-\frac{\alpha}{r}\rt)y={\cal E}y
\eeq
where $\alpha=Z,~{\cal E}=m\omega_L+2(|m|+1)\tilde{\omega}_r-\epsilon $  and
\beq
{\cal H}=\frac{d^2}{dr^2}+\lt(-2\tilde{\omega}_r r+\frac{2|m|+1}{r}\rt)\frac{d}{dr}.
\eeq

Applying the St\"ackel transform, we obtain
\beqa
{\cal H}'y&=&Z y,\nonumber\\
{\cal H}'&=&r(H-{\cal E})\nonumber\\
  &=&r\frac{d^2}{dr^2}+[-2\omega_L r^2+2|m|+1]\frac{d}{dr}+[\epsilon-m\omega_L-2(|m|+1)\tilde{\omega}_r] r.
\eeqa
The St\"ackel transformed Hamiltonian ${\cal H}'$ allows for a $sl(2)$ algebraization if
\beq
\epsilon-m\omega_L-2(|m|+1)\tilde{\omega}_r\equiv c_1=-n[(n-1)a_3+b_2]\equiv 2\tilde{\omega}_r n
\eeq
which gives
\beq
\epsilon=m\omega_L+2(n+|m|+1)\tilde{\omega}_r,~~~~n=0,1,2,\cdots
\eeq
Indeed, for such $\epsilon$ values, ${\cal H}'$ is dependent on integer parameter $n$ and can be expressed in terms of the $sl(2)$ generators as
\beq
{\cal H}'=J^0J^-+2\tilde{\omega}_r J^++\lt[\frac{n}{2}+2|m|+1\rt]J^-.
\eeq

This is one of our main result in this section. The Schr\"odinger equation  ${\cal H}'y=Z y$ can now be analytically solved and solutions are given by polynomials. The solutions to the original system are then obtained by CCM. 

In the following, we present the explicit expressions for solutions corresponding to the $n=1, 2$ cases.  For $n=1$, the eigenvalues and eigenfunctions of $\mathcal{H}'y=Z y$ are given by
\begin{align}
   Z=\pm \sqrt{2\tilde{\omega}_r(2|m|+1)},  \qquad 
    y=r \pm \sqrt{\frac{|m|+1/2}{\tilde{\omega}_r}}. 
\end{align}
It follows by CCM that the energy and wavefunctions for the original system are 
\begin{align}
    &\begin{aligned}
    E_1=m\omega_L+2(|m|+2)\tilde{\omega}_r+\eta,
    \end{aligned}\\
    &\begin{aligned}
        \phi_{1\pm}({\bf r})=\frac{r^{|m|}}{\sqrt{2\pi}}\exp\left(im\theta-\frac{\tilde{\omega}_r}{2}r^2\right)\left(r \pm \sqrt{\frac{|m|+1/2}{\tilde{\omega}_r}}\right), 
    \end{aligned}
\end{align}
corresponding to the two values of the model parameter $Z$, respectively.
For $n=2$, we have 
\begin{align}
    &\begin{aligned}
        Z_1=0,\quad Z_{2,3}=\pm 2\sqrt{(3+4|m|)\tilde{\omega}_r}, 
    \end{aligned}\\
    &\begin{aligned}
        y_1=r^2-\frac{1+m}{\tilde{\omega}_r},\quad y_{2,3}=r^2 \pm \sqrt{\frac{3+4m}{\tilde{\omega}_r}} r+\frac{1+2m}{2 \tilde{\omega}_r}. 
    \end{aligned}
\end{align}
By CCM, the corresponding energy and the wavefunctions of the original system are
\begin{align}
    &\begin{aligned}
    E_2=m\omega_L+2(m+3)\tilde{\omega}_r+\eta,
    \end{aligned}\\
    &\begin{aligned}
        &\phi_{2(1)}({\bf r})=\frac{r^{|m|}}{\sqrt{2\pi}}\exp\left(im\theta-\frac{\tilde{\omega}_r}{2}r^2\right)\left(r^2-\frac{1+m}{\tilde{\omega}_r}\right),\\    
        &\phi_{2(2,3)}({\bf r})=\frac{r^{|m|}}{\sqrt{2\pi}}\exp\left(im\theta-\frac{\tilde{\omega}_r}{2}r^2\right)\left(r^2 \pm \sqrt{\frac{3+4m}{\tilde{\omega}_r}} r+\frac{1+2m}{2 \tilde{\omega}_r}\right). 
    \end{aligned}
\end{align}
There are 3 independent wavefunctions for $n=2$, as requied by the $sl(2)$ symmetry.
Note that $Z=0$ is not physically interesting case because it means that there is no Coulomb interaction.

\section{Two Coulombically repelling electrons on a sphere}\label{two-electron-on-sphere}

In this section, we study a system with two electrons trapped on a sphere \cite{ezra1982}. 
This model was shown to be QES in \cite{loos2009} and analytic expression for its energy spectrum was also given. In \cite{zhang2012}, general closed-form expressions for both the energy spectrum and wavefunctions were derived by means of the Bethe ansatz method. Here, we find the hidden $sl(2)$ symmetry and analytic solutions by applying the St\"ackel transform and CCM approach described in section 2.

Consider a system of two electrons, interacting via a Coulomb potential, but constrained to remain on the surface of a sphere of radius $R$. The Hamiltonian of the system (in atomic units) is \cite{loos2009}
\beq
H_0=-\frac{1}{2}\left(\nabla_1^2+\nabla_2^2\right)-\frac{1}{u},
\eeq
where $u=|{\bf r}_1-{\bf r}_2|$ is the inter-electronic distance. The Schr\"odinger wave function of the system can be
separated into a product of spin, angular and inter-electron components, with the inter-electron wave function $\Psi(u)$ satisfying the ODE \cite{loos2009}
\beq
\left(\frac{u^2}{4R^2}-1\right)\frac{d^2\Psi}{du^2}+\left(\frac{\delta u}{4R^2}-\frac{1}{\gamma u}\right)\frac{d\Psi}{du}
   +\frac{\Psi}{u}=E\Psi,
\eeq
where $\delta$ and $\gamma$ are certain parameters. Introduce dimensionless variable
$z=\frac{u}{2R}$. Then the above ODE can be written as
\beq
{\cal H}\Psi=\left[H-\frac{\alpha}{z}\right]\Psi={\cal E}\Psi
\eeq
where $\alpha=-2R$, ${\cal E}=4R^2E$ and
\beq
H=(z^2-1)\frac{d^2}{dz^2}+\lt(\delta z-\frac{1/\gamma}{z}\rt)\frac{d}{dz}
\eeq

Applying the St\"ackel transform, we get
\beqa
&&{\cal H}'\Psi=-2R\Psi,\nonumber\\
&&{\cal H}'=z(H-{\cal E})=z(z^2-1)\frac{d^2}{dz^2}+\lt(\delta z^2-\frac{1}{\gamma}\rt)\frac{d}{dz}-4R^2E z.
\eeqa
Then the St\"ackel transformed Hamiltonian ${\cal H}'$ allows for an $sl(2)$ algebraization if
\beq
-4R^2E\equiv c_1=-n[(n-1)a_3+b_2]\equiv -n[n-1+\delta]
\eeq
which gives the exact energies of the system obtained in \cite{loos2009,zhang2012}
\beq
E=\frac{1}{4R^2}n(n-1+\delta),~~~~~n=0,1,2,\cdots.
\eeq
Indeed, for such $E$ values, ${\cal H}'$ is dependent on integer parameter $n$ and can be expressed in terms of the $sl(2)$ generators as
\beq
{\cal H}'=-J^+J^0-J^0J^--\left[\frac{(3n-2)}{2}+\delta\right]J^+ -\left(\frac{1}{\gamma}+\frac{n}{2}\right)J^-.
\eeq
This provides an $sl(2)$ algebraization of the St\"ackel equivalent of the two-electron system. Solving this system algebraically and using CCM, we can obtain closed-form expressions for analytical solutions of the original system. As examples, in the following, we present the results for the $n=1,2$ cases.

For $n=1$, the eigenvalues and eigenfunctions of $\mathcal{H}'$ are
\begin{align}
   R=\pm \frac{1}{2} \sqrt{\frac{\delta}{\gamma}}, \qquad
    y=z \pm \sqrt{\frac{1}{\gamma \delta}}. 
\end{align}
By CCM, the energy and the wavefunctions for the original system are
\begin{align}
    E_1=\gamma,\qquad
    \Psi_{1\pm}(u)= \pm \frac{1+\gamma u}{\sqrt{\gamma \delta}}. 
\end{align}
Sumilarly for $n=2$, we have 
\begin{align}
    &\begin{aligned}
        R_1=0,\quad R_{2,3}=\pm \sqrt{\frac{3+2\gamma+2\delta+\delta \gamma}{2\gamma}}, 
    \end{aligned}\\
    &\begin{aligned}
        y_1=z^2-\frac{1+\gamma}{\gamma(1+\delta)},\quad y_{2,3}=z^2 \pm \frac{1}{2+\delta} \sqrt{\frac{2(3+2\gamma+2\delta+\gamma\delta)}{\gamma}} z+\frac{1}{\gamma (2+\delta)}. 
    \end{aligned}
\end{align}
There are 3 independent eigenfunctions for $n=2$, in agreement with the requirement of the hidden $sl(2)$ symmetry. However, from physical perspective, the radius of the sphere $R$ cannot be 0. Thus, the solution $R_1=0$ above is nonphysical and will be discarded. By CCM, we obtain the energy and 2 physical 2 wavefunctions of original system
\begin{align}
    &\begin{aligned}
        E_2=\frac{\gamma(1+\delta)}{3+2\delta+\gamma(2+\delta)},
    \end{aligned}\\
    &\begin{aligned}
        \Psi_{2(2)}(u)=\Psi_{2(3)}(u)=\frac{1+\gamma u}{\gamma(2+\delta)}+\frac{\gamma u^2}{6+4\delta+2\gamma(2+\delta)} 
    \end{aligned}
\end{align}
associated with the 2 values $R_{2,3}$ of the model parameter (the radius) $R$, respectively.

\section{Inverse quartic power potential}

In this section, we consider a model with the following quartic inverse power potential with a strongly singular repulsive core at the origin \cite{predazzi1962} 
\beq
H_0=-\frac{1}{2}\frac{d^2}{dr^2}+V(r),\qquad V(r)=\frac{a}{r^4}+\frac{b}{r^3}+\frac{c}{r^2}+\frac{d}{r}, \quad d > 0.
\eeq
Models with singular potentials have applications in real-word physics. For example, the Mie-type potential and the Lennard-Jones potential describe molecular vibrations \cite{ghanbari2020} and molecular simulations \cite{wang2020}, respectively.  

If the model parameters $a, b, c d$ satisfy certain constraints, the model with the above potential is QES and the corresponding closed-form expressions for energies and wave functions have been derived in \cite{agboola2013} by means of the Bethe ansatz method \cite{zhang2012}. A $sl(2)$ algebraization for a special inverse quartic power potential was obtained in \cite{shifman1989}. Here we show the hidden $sl(2)$ algebra symmetry for the model with the above general inverse quartic potential by applying the St\"ackel transform. We also obtain its analytic solutions via the CCM method. 

The Schr\"odinger equation of the system $H_0\psi(r)=E\psi(r)$ can be written as \cite{agboola2013}
\begin{equation}\label{quartic inverse radial schrodinger equation}
   \left[-\frac{d^2}{dr^2}+\frac{2a}{r}+\frac{2b}{r^2}+\frac{2c}{r^3}+\frac{2d}{r^4}\right]\psi(r)=2E\psi(r).
\end{equation}
Making the gauge transformation
\begin{equation}\label{quartic inverse wavefunction}
    \psi(r)=\exp \left[\left(1+\frac{c}{\sqrt{2d}}\right)\ln r+Br-\frac{\sqrt{2d}}{r}\right]f(r). 
\end{equation}
The Schr\"odinger equation (\ref{quartic inverse radial schrodinger equation} becomes \footnote{Note that there are some typos in (2.5) of \cite{agboola2013}, which are corrected here.}
\begin{equation}\label{quartic inverse ODE}
\begin{split}
        &\frac{d^2}{dr^2}f(r)+2\left(B+\frac{1+c/\sqrt{2d}}{r}+\frac{\sqrt{2d}}{r^2}\right)\frac{d}{dr}f(r)\\
        &\quad+\left(B^2+2E+\frac{2B (1+c/\sqrt{2d})-2a}{r}+\frac{(1+c/\sqrt{2d})\,c/\sqrt{2d}-2b+2B\sqrt{2d}}{r^2}\right)f(r)=0. 
\end{split}
\end{equation}
By means of CCM, (\ref{quartic inverse ODE}) can be expressed as 
\begin{equation}\label{quartic inverse alpha varepsilon H}
    \begin{split}
        &\mathcal{H}=\left(H-\frac{\alpha}{r^2}\right)f(r)=\varepsilon f(r),\\
        &\alpha=-\frac{c}{\sqrt{2d}}\left(1+\frac{c}{\sqrt{2d}}\right)+2b-2B\sqrt{2d},\\
        &\varepsilon=-(2E+B^2),\\
        &H=\frac{d^2}{dr^2}+2\left(B+\frac{1+c/\sqrt{2d}}{r}+\frac{\sqrt{2d}}{r^2}\right)\frac{d}{dr}+\frac{2B (1+c/\sqrt{2d})-2a}{r}. 
    \end{split}
\end{equation}

Using the St\"ackel transformation, we have
\begin{equation}\label{quartic inverse new operator H}
\begin{split}
        \mathcal{H}'f(r)=&\alpha f(r),\\
        \mathcal{H}'=&r^2(H-\varepsilon)=r^2\frac{d^2}{dr^2} +2\left[Br^2+\left(1+\frac{c}{\sqrt{2d}}\right)r+\sqrt{2d}\right] \frac{d}{dr}\\    
        &\qquad\qquad\qquad+(2E+B^2)r^2+2\left(B-a+B\frac{c}{\sqrt{2d}}\right)r.
\end{split}
\end{equation}
$\mathcal{H}'$ has a hidden $sl(2)$ algebraic structure if
\begin{eqnarray}
    2\left(B-a+B\frac{c}{\sqrt{2d}}\right)=-2nB,\qquad\quad 2E+B^2=0,
\end{eqnarray}
which give the energy spectrum of the system,
\begin{align}\label{quartic inverse energy n}
    \begin{aligned}
        E=-\frac{1}{2}B^2, \qquad B=\frac{a}{n+1+\frac{c}{\sqrt{2d}}}.
    \end{aligned}
\end{align}
Indeed for such $E$ and $B$ values given above, $\mathcal{H}'$ can be expressed in terms of the $sl(2)$ differential operators as follows: 
\begin{equation}
\begin{split}
        \mathcal{H}'=J^0J^0-\frac{2a}{n+1+\frac{c}{\sqrt{2d}}}J^++\left(n+1+\frac{2c}{\sqrt{2d}}\right)J^0+2\sqrt{2d}\,J^-
        +\frac{n}{4}\left(n+2+\frac{4c}{\sqrt{2d}}\right). 
\end{split}
\end{equation}

For $n=1$, the eigenvalues, eigenfunctions and constraints of model parameters for $\mathcal{H}'$ are
\begin{align}
    &\begin{aligned}
        \alpha=1+\frac{c}{\sqrt{2d}} \pm \sqrt{\left(1+\frac{c}{\sqrt{2d}}\right)^2- \frac{4a\sqrt{2d}}{2+\frac{c}{\sqrt{2d}}}},
    \end{aligned}\\
    &\begin{aligned}
        f(r)=r-\frac{2+\frac{c}{\sqrt{2d}}}{2a}\left[-1-\frac{c}{\sqrt{2d}} \pm \sqrt{\left(1+\frac{c}{\sqrt{2d}}\right)^2- \frac{4a\sqrt{2d}}{2+\frac{c}{\sqrt{2d}}}}\,\right],
     \end{aligned}\\
     &\begin{aligned}
     b=\frac{1}{2}\left[\alpha+\frac{c}{\sqrt{2d}}\left(1+\frac{c}{\sqrt{2d}}\right)\right]+\frac{a\sqrt{2d}}{2+\frac{c}{\sqrt{2d}}}.
     \end{aligned}
\end{align}
By CCM, we obtain the corresponding energy and wavefunctions of the original system for $n=1$
\begin{align}\label{quartic inverse energy 1}
    &\begin{aligned}
        E_1=-\frac{1}{2}\frac{a^2}{\left(2+\frac{c}{\sqrt{2d}}\right)^2},
    \end{aligned}\\
    &\begin{aligned}\label{quartic inverse wavefunction 1}
        \psi_1(r)=\exp \left[\left(1+\frac{c}{\sqrt{2d}}\right)\ln r+\frac{a}{2+\frac{c}{\sqrt{2d}}}\,r-\frac{\sqrt{2d}}{r}\right]f(r),
    \end{aligned}
\end{align}
corresponding to the two different values of model parameter $\alpha$ given above.

For $n=2$, eigenvalues and constraints of the model parameters for $\mathcal{H}'$ are given by 
\begin{equation}\label{quartic inverse eigenvalue function 2}
    \alpha^3-\left(5+\frac{4c}{\sqrt{2d}}\right)\alpha^2+\left(6+\frac{2c^2}{d}+\frac{10c}{\sqrt{2d}}-\frac{16a\sqrt{2d}}{3+\frac{c}{\sqrt{2d}}}\right)\alpha+\frac{8a(2c+3\sqrt{2d})}{3+\frac{c}{\sqrt{2d}}}=0, 
\end{equation}
\beq
b=\frac{1}{2}\left[\alpha+\frac{c}{\sqrt{2d}}\left(1+\frac{c}{\sqrt{2d}}\right)\right]+\frac{a\sqrt{2d}}{3+\frac{c}{\sqrt{2d}}}.
\eeq
There are 3 independent eigenvalues in (\ref{quartic inverse eigenvalue function 2}), which give 3 independent eigenfunctions for $n=2$, as expected from the hidden $sl(2)$ algebra structure of $\mathcal{H}'$. The energy of the original system for $n=2$ is
\begin{equation}\label{quartic inverse energy 2}
    E_2=-\frac{1}{2}\frac{a^2}{\left(3+\frac{c}{\sqrt{2d}}\right)^2}. 
\end{equation}

\section{Inverse sextic power potential}
In this section, we consider the inverse sextic power potential, which was used to study unrenormalizable interaction in field theory \cite{pais1964}.
\begin{equation}
    V(r)=\frac{c}{r^4}+\frac{d}{r^6},\quad d>0. 
\end{equation}
Analytic solutions of the model were studied by different methods\cite{znojil1990,agboola2013}. We apply St\"ackel transform and CCM method to provide its hidden $sl(2)$ symmetry and obtain wavefucntions and energies analytically. 
The radial Schr\"odinger equation is 
\begin{eqnarray}\label{inverse sextic schrodinger equation}
    \left[-\frac{d^2}{dr^2}+\frac{l(l+1)}{r^2}+\omega^2 r^2+\frac{2c}{r^4}+\frac{2d}{r^6}\right]\psi(r)=2E\psi(r). 
\end{eqnarray}
Making the gauge transformation, we set a wavefunction 
\begin{equation}\label{inverse sextic wavefunction}
    \psi(r)=\exp\left[-\frac{\omega}{2}r^2-\frac{\sqrt{2d}}{2}r^2+\left(\frac{3}{2}+\frac{c}{\sqrt{2d}}\right)\ln r\right]f(r). 
\end{equation}
Substitute (\ref{inverse sextic wavefunction}) into (\ref{inverse sextic schrodinger equation}) and change variable $z=r^2$, the Schr\"odinger equation transforms to 
\begin{equation}\label{inverse sextic ODE}
\begin{split}
        &z\frac{d^2}{dz^2}f(z)+\left(\frac{c}{\sqrt{2d}}+2+\frac{\sqrt{2d}}{z}-\omega z\right)\frac{d}{dz}f(z)\\
        &+\left[\frac{1}{2}E-\left(\frac{c}{2\sqrt{2d}}+1\right)+\frac{1}{z}\left(\frac{c^2}{8d}+\frac{c}{2\sqrt{2d}}+\frac{3}{16}-\frac{l(l+1)}{4}+\frac{\sqrt{2d}\omega}{2}\right)\right]f(z)=0. 
\end{split}
\end{equation}
Using CCM, (\ref{inverse sextic ODE}) can be written as 
\begin{equation}\label{inverse sextic alpha epsilon H}
\begin{split}
        &\mathcal{H}=\left(H-\frac{\alpha}{z}\right)f(z)=\varepsilon f(z),\\
        &\alpha=-\left(\frac{c^2}{8d}+\frac{c}{2\sqrt{2d}}+\frac{3}{16}-\frac{l(l+1)}{4}+\frac{\sqrt{2d}\omega}{2}\right)\\
        &\varepsilon=\left(\frac{c}{2\sqrt{2d}}+1\right)-\frac{1}{2}E\\
        &H=z\frac{d^2}{dz^2}+\left(\frac{c}{\sqrt{2d}}+2+\frac{\sqrt{2d}}{z}-\omega z\right)\frac{d}{dz}. 
\end{split}
\end{equation}
Applying St\"ackel transformation, we derive
\begin{equation}\label{inverse sextic operator H}
    \begin{split}
        &\mathcal{H}'f(z)=\alpha f(z),\\
        &\mathcal{H}'=z^2\frac{d^2}{dz^2}+\left[\left(\frac{c}{\sqrt{2d}}+2\right)z+\sqrt{2d}-\omega z^2\right]\frac{d}{dz}+\left[\frac{1}{2}E-\left(\frac{c}{2\sqrt{2d}}+1\right)\right]z.
    \end{split}
\end{equation}
$\mathcal{H}'$ has $sl(2)$ algebraization if
\begin{equation}
    \frac{1}{2}E-\left(\frac{c}{2\sqrt{2d}}+1\right)=\omega n,
\end{equation}
which gives the energy of the system obtained in \cite{agboola2013},
\begin{equation}\label{inverse sextic energy}
    E=\left(2n+2+\frac{c}{\sqrt{2d}}\right)\omega. 
\end{equation}
With $E$ given in (\ref{inverse sextic energy}), $\mathcal{H}'$ can be writtes in terms of $sl(2)$ differential operators
\begin{equation}
\mathcal{H}'=J^0J^0+\omega\sqrt{d}J^++\left(n+1+\frac{c}{\sqrt{2d}}\right)J^0+\sqrt{2d}J^-+\frac{1}{4}n\left(n+2+\frac{2c}{\sqrt{2d}}\right). 
\end{equation}

For $n=1$, eigenvalues and eigenfunctions of $\mathcal{H}'$ are given by \begin{align}
    &\begin{aligned}
        \alpha=1+\frac{c}{2\sqrt{2d}}\pm\sqrt{\left(1+\frac{c}{2\sqrt{2d}}\right)^2+\omega\sqrt{2d}}
    \end{aligned}\\
    &\begin{aligned}
        f(z)=z-\frac{1}{\omega}\left(1+\frac{c}{2\sqrt{2d}}\pm \sqrt{\left(1+\frac{c}{2\sqrt{2d}}\right)^2+\omega\sqrt{2d}}\right)
    \end{aligned}
\end{align} 
and the constraints of the model parameters are
\begin{equation}
    \begin{split}
\left(\frac{c^2}{8d}-\frac{13}{16}-\frac{l(l+1)}{4}+\frac{\sqrt{2d}\omega}{2}\right)\pm\sqrt{\left(1+\frac{c}{2\sqrt{2d}}\right)^2+\omega\sqrt{2d}}=0.
    \end{split}
\end{equation}
Via CCM, we obtain the the energy and wavefunction of original system for $n=1$
\begin{align}
    &\begin{aligned}
        E_1=\left(4+\frac{c}{\sqrt{2d}}\right)\omega,
    \end{aligned}\\
    &\begin{aligned}
       \psi_1(r)=&\exp\left[-\frac{\omega}{2}r^2-\frac{\sqrt{2d}}{2}r^2+\left(\frac{3}{2}+\frac{c}{\sqrt{2d}}\right)\ln r\right]\\
       &\cdot\left[r^2-\frac{1}{\omega}\left(1+\frac{c}{2\sqrt{2d}}\pm \sqrt{\left(1+\frac{c}{2\sqrt{2d}}\right)^2+\omega\sqrt{2d}}\right)\right]
    \end{aligned}
\end{align}
For $n=2$, the eigenvalues and the constraints for the model parameters are determined by
\begin{equation}
    \alpha+\left(\frac{c^2}{8d}+\frac{c}{2\sqrt{2d}}+\frac{3}{16}-\frac{l(l+1)}{4}+\frac{\sqrt{2d}\omega}{2}\right)=0,
\end{equation}
\begin{equation}
    \alpha^3-\left(8+\frac{3c}{\sqrt{2d}}\right)\alpha^2+\left(\frac{c^2}{d}+\frac{10c}{\sqrt{2d}}-4\omega\sqrt{2d}+12\right)\alpha+4\omega(c+3\sqrt{2d})=0,
\end{equation}
respectively.
There are 3 independent eigenfunctions for $n=2$, as expected from the hidden $sl(2)$ symmetry of $\mathcal{H}'$. The energy of original system for $n=2$ is 
\begin{equation}
    E_2=\left(6+\frac{c}{\sqrt{2d}}\right)\omega. 
\end{equation}

\section{Quantum Newtonian cosmology}
In this section, we apply our procedure to Newtonian cosmology \cite{mccrea1934, mccrea1955}. A quantum Newtonian cosmology model has recently been studied in \cite{vieira2015,vieira2019} using the biconfluent Heun functions. We discover the hidden $sl(2)$ algebra symmetry of the model and obtain its corresponding $sl(2)$ algebraization via St\"ackel transform and CCM approach.

The effective potential of a particle moving in the Newtonian universe is \cite{vieira2019}
\begin{equation}\label{Newtonian effective potential}
    V_{eff}(r)=-\frac{4 \pi G \mu}{3} \left[ A_d+A_q r+\left( A_v+ \frac{\Lambda}{8 \pi G}r^2+\frac{A_m}{r}+\frac{A_r}{r^2} \right) \right].
\end{equation}

The corresponding Schr\"odinger equation $H_0\psi(r)=E\psi(r)$, where $H_0=-\frac{\hbar^2}{2\mu}\frac{d^2}{dr^2}+V_{eff}(r)$, can be written as the form,

\begin{equation}\label{Newtonian Schrodinger equation}
    \frac{d^2}{d r^2} \psi(r)+\left(B_1+B_2 r+B_3 r^2+\frac{B_4}{r}+\frac{B_5}{r^2}\right)\psi(r)=0,
\end{equation}
where the parameters $B_1, B_2, B_3, B_4$ and $B_5$ are
\begin{equation}\label{Newtonian parameters}
    \begin{split}
        &B_1=\frac{2\mu E}{\hbar^2}+\frac{8\pi G\mu^2}{3\hbar^2}A_d, \qquad B_2=\frac{8\pi G\mu^2}{3\hbar^2}A_q,\\
        &B_3=\frac{8\pi G\mu^2}{3\hbar^2} \left(A_v+\frac{\Lambda}{8\pi G}\right), \qquad B_4=\frac{8\pi G\mu^2}{3\hbar^2} A_m, \qquad B_5=\frac{8\pi G\mu^2}{3\hbar^2} A_r. 
    \end{split}
\end{equation}

In terms of new variable $x=\tau\, r$ with $\tau=(-B_3)^{1/4}$, the Schr\"odinger equation can be expressed as the form
\begin{equation}\label{Newtonian Schrodinger equation x}
    \frac{d^2}{d x^2}\psi(x)+\left(\frac{B_1}{\tau^2}+\frac{B_2}{\tau^3}\, x-x^2+\frac{B_4/\tau}{x}+\frac{B_5}{x^2}\right) \psi(x)=0.
\end{equation}

After making the gauge transformation 
\begin{equation}\label{Newtonian solution}
    \psi(x)=x^{\frac{1}{2}(1-\sqrt{1-4 B_5})}\,\exp\left(-\frac{1}{2}x^2+\frac{B_2}{2 \tau^3}x\right)\, f(x),
\end{equation}
the ODE  (\ref{Newtonian Schrodinger equation x}) becomes
\begin{eqnarray}\label{Newtonian ODE}
    &&\frac{d^2}{dx^2}f(x)+\left( \frac{1-\sqrt{1-4 B_5}}{x}-2x+\frac{B_2}{\tau^3}\right)\frac{d}{dx}f(x)\nonumber\\
    &&\quad+\left( \frac{B_2(1-\sqrt{1-4 B_5})/(2\tau^3)+B_4/\tau}{ x} +\sqrt{1-4 B_5}-2+\frac{B_1}{\tau^2}+\frac{B_2^2}{4 \tau^6}\right) f(x)=0.
\end{eqnarray}
This ODE can be rewritten as 
\begin{equation}\label{Newtonian alpha varepsilon H}
\begin{split}
    &\mathcal{H}=\left(H-\frac{\alpha}{x}\right)f(x)=\varepsilon f(x),\\
    &\alpha=- \frac{ B_2(1-\sqrt{1-4 B_5})}{2\tau^3}-\frac{B_4}{\tau},\\
    &\varepsilon=2-\sqrt{1-4 B_5}-\frac{B_1}{\tau^2}-\frac{B_2^2}{ 4\tau^6},\\
    &H=\frac{d^2}{dx^2}+\left( \frac{1-\sqrt{1-4 B_5}}{x}-2x+\frac{B_2}{\tau^3}\right)\frac{d}{dx}.
\end{split}
\end{equation}
Applying the St\"ackel transform, we obtain 
\begin{equation}\label{Newtonian new operator H}
    \begin{split}
      \mathcal{H}'& f(x)=\alpha f(x),\\
      \mathcal{H}'&=x(H-\varepsilon)\\
        &=x \frac{d^2}{dx^2}+\left(1-\sqrt{1-4 B_5}-2x^2+\frac{B_2}{\tau^3} x\right)\frac{d}{dx}+\left( \sqrt{1-4 B_5}-2+\frac{B_1}{\tau^2}+\frac{B_2^2}{4 \tau^6}\right) x.
    \end{split}
\end{equation}
Then $\mathcal{H}'$ has a $sl(2)$ algebraization if 
\begin{equation}\label{Newtonian parameter constrain}
     \sqrt{1-4 B_5}-2+\frac{B_1}{\tau^2}+\frac{B_2^2}{4 \tau^6}=2 n, 
\end{equation}
which gives  
\begin{equation}\label{Newtonian energy}
    E=\left[2(n+1)-\sqrt{1-4 B_5}\right]\,\frac{\hbar^2\,\tau^2}{2\mu}-\frac{\hbar^2}{8\mu\tau^4}B_2^2-\frac{4\pi \mu G}{3}A_d. 
\end{equation}
Indeed, if $E$ is given by the above formula, $\mathcal{H}'$ can be expressed in terms of the $sl(2)$ differential operators as 
\begin{equation}\label{Newtonian sl(2)}
    \mathcal{H}'=J^0 J^-+2J^++\frac{B_2}{\tau^3}J^0+ \left(1-\sqrt{1-4 B_5}+\frac{n}{2} \right)J^-+\frac{B_2}{2 \tau^3}\,n. 
\end{equation}

For $n=1$, the eigenvalues and the corresponding eigenfunctions for $\mathcal{H}'$ are given by
\begin{align}
    &\begin{aligned}
        \alpha=\frac{B_2}{2\tau^3} \pm \sqrt{\frac{B_2^2}{4\tau^6}+2(1-\sqrt{1-4B_5})},
    \end{aligned}\\
    &\begin{aligned}
        f(x)=x-\frac{B_2}{4\tau^3}\pm\sqrt{\frac{B_2^2}{16\tau^6}+\frac{1-\sqrt{1-4B_5}}{2}},
    \end{aligned}
\end{align}
where the model parameters obey the following constrains
\begin{align}
    \begin{aligned}
       2\tau^2\, B_4=B_2(-2+\sqrt{1-4 B_5}) \pm \sqrt{B_2^2-8(-1+\sqrt{1-4 B_5})\tau^6}.
    \end{aligned}
\end{align}
By CCM, the energy and the wavefunctions of the initial system for $n=1$ are
\begin{align}
    &\begin{aligned}
       E=\left[4-\sqrt{1-4 B_5}\right]\,\frac{\hbar^2\,\tau^2}{2\mu}-\frac{\hbar^2}{8\mu\tau^4}B_2^2-\frac{4\pi \mu G}{3}A_d, 
    \end{aligned}\\
    &\begin{aligned}
        \psi_{1\pm}(r)=&[(-B_3)^{1/4} r]^{\frac{1}{2}(1-\sqrt{1-4 B_5})}\,\exp\left(\frac{B_2 r}{2\sqrt{-B_3}}-\frac{1}{2}\sqrt{-B_3} r^2\right)\\
        &\times(-B_3)^{1/4}\left[r+\frac{B_2 \pm \sqrt{B_2^2-8(-B_3)^{3/2}(-1+\sqrt{1-4B5})}}{4B_3}\,\right],
    \end{aligned}
\end{align}
corresponding to two different values of the model parameter $\alpha$ given above.

For $n=2$, the eigenvalues and constraints of model parameters for $\mathcal{H}'$ can be determined from the algebraic equations,
\begin{align}
    &\begin{aligned}\label{Newtonian cosmology parameter constrain n=2}
        \alpha^3-3B_2\alpha^2+(2B_2^2-12\tau^6+8\sqrt{1-4B_5}\tau^6)\alpha+8B_2\tau^6(1-\sqrt{1-4B_5})=0,
    \end{aligned}\\
    &\begin{aligned}
    2\tau^3\,\alpha+ B_2(1-\sqrt{1-4 B_5})+2\tau^2\,B_4=0.
\end{aligned}
\end{align}
The cubic equation (\ref{Newtonian cosmology parameter constrain n=2}) indicates that there are 3 independent eigenvalues. Correspondingly there are 3 different wavefunctions for $n=2$, as expected from the $sl(2)$ algebraic structure of $\mathcal{H}'$. By CCM, we can obtain the energy of the original system,
\beq
E=\left[6-\sqrt{1-4 B_5}\right]\,\frac{\hbar^2\,\tau^2}{2\mu}-\frac{\hbar^2}{8\mu\tau^4}B_2^2-\frac{4\pi \mu G}{3}A_d
\eeq
and the corresponding wavefunctions whose explicit expressions will be omitted here due to their long and complicated forms.

\section{Conclusion}

One of the main results of this paper is the development of a new method based on the St\"ackel transform, which brings a non-Lie algebraic QES Hamiltonian into its St\"ackel equivalent that has a $sl(2)$ algebraization.   This makes the original system "amenable" in algebraic form in the Lie $sl(2)$ algebra setting, and allows us to determine the energy spectrum and analytical solutions of the original system by means of the approach of CCM.
We apply this approach to a wide range of QES models of relevance for physical applications, whose original (gauge-transformed) Hamiltonians do not possess any $sl(2)$ algebraizations. Based on $sl(2)$ algebraizations and solutions of the St\"ackel equivalent systems, we also obtain the energy spectrum and analytical wavefunctions of the original systems via CCM. In each case, we present their explicit expressions for $n=1,2$.

 
Our results show that the St\"ackel transform and CCM are applicable to a wide set of problems. It is interesting to generalize the procedure to systems associated with higher-rank Lie algebras such as $sl(m+1)$. This is under investigation and results will be presented elsewhere. 

\section*{Acknowledgement}
IM was supported by by Australian Research Council Future Fellowship FT180100099, and YZZ was supported by Australian Research Council Discovery Project DP190101529.

\bibliographystyle{IEEEtran}
\bibliography{Reference.bib}

\end{document}